\documentclass[a4paper,11pt]{article}

\makeatletter

\gdef\@fpheader{\newline}
\gdef\@journal{jhep}

\RequirePackage{amsmath}
\RequirePackage{amssymb}
\RequirePackage{epsfig}
\RequirePackage{CJK}
\RequirePackage{graphicx}
\RequirePackage[numbers,sort&compress]{natbib}
\RequirePackage{color}
\RequirePackage[colorlinks=true
,urlcolor=blue
,anchorcolor=blue
,citecolor=blue
,filecolor=blue
,linkcolor=blue
,menucolor=blue
,pagecolor=blue
,linktocpage=true
,pdfproducer=medialab
,pdfa=true
,CJKbookmarks
]{hyperref}

\newif\ifnotoc\notocfalse
\newif\ifemailadd\emailaddfalse
\newif\iftoccontinuous\toccontinuousfalse

\def\@subheader{\@empty}
\def\@keywords{\@empty}
\def\@abstract{\@empty}
\def\@xtum{\@empty}
\def\@dedicated{\@empty}
\def\@arxivnumber{\@empty}
\def\@collaboration{\@empty}
\def\@collaborationImg{\@empty}
\def\@proceeding{\@empty}
\def\@preprint{\@empty}

\newcommand{\subheader}[1]{\gdef\@subheader{#1}}
\newcommand{\keywords}[1]{\if!\@keywords!\gdef\@keywords{#1}\else%
\PackageWarningNoLine{\jname}{Keywords already defined.\MessageBreak Ignoring last definition.}\fi}
\renewcommand{\abstract}[1]{\gdef\@abstract{#1}}
\newcommand{\dedicated}[1]{\gdef\@dedicated{#1}}
\newcommand{\arxivnumber}[1]{\gdef\@arxivnumber{#1}}
\newcommand{\proceeding}[1]{\gdef\@proceeding{#1}}
\newcommand{\xtumfont}[1]{\textsc{#1}}
\newcommand{\correctionref}[3]{\gdef\@xtum{\xtumfont{#1} \href{#2}{#3}}}
\newcommand\jname{JHEP}
\newcommand\acknowledgments{\section*{Acknowledgments}}

\newcommand\preprint[1]{\gdef\@preprint{\hfill #1}}

\newcommand\note[2][]{%
\if!#1!%
\stepcounter{footnote}\footnotetext{#2}%
\else%
{\renewcommand\thefootnote{#1}%
\footnotetext{#2}}%
\fi}

\newtoks\auth@toks
\renewcommand{\author}[2][]{%
  \if!#1!%
    \auth@toks=\expandafter{\the\auth@toks#2\ }%
  \else
    \auth@toks=\expandafter{\the\auth@toks#2$^{#1}$\ }%
  \fi
}

\newtoks\affil@toks\newif\ifaffil\affilfalse
\newcommand{\affiliation}[2][]{%
\affiltrue
  \if!#1!%
    \affil@toks=\expandafter{\the\affil@toks{\item[]#2}}%
  \else
    \affil@toks=\expandafter{\the\affil@toks{\item[$^{#1}$]#2}}%
  \fi
}

\newtoks\email@toks\newcounter{email@counter}%
\newcommand{\emailAdd}[1]{%
\emailaddtrue%
\ifnum\theemail@counter>0\email@toks=\expandafter{\the\email@toks, \@email{#1}}%
\else\email@toks=\expandafter{\the\email@toks\@email{#1}}%
\fi\stepcounter{email@counter}}
\newcommand{\@email}[1]{\href{mailto:#1}{\tt #1}}

\newcommand*\collaboration[1]{\gdef\@collaboration{#1}}
\newcommand*\collaborationImg[2][]{\gdef\@collaborationImg{#2}}

\newcommand\afterLogoSpace{\smallskip}
\newcommand\afterSubheaderSpace{\vskip3pt plus 2pt minus 1pt}
\newcommand\afterProceedingsSpace{\vskip21pt plus0.4fil minus15pt}
\newcommand\afterTitleSpace{\vskip23pt plus0.06fil minus13pt}
\newcommand\afterRuleSpace{\vskip23pt plus0.06fil minus13pt}
\newcommand\afterCollaborationSpace{\vskip3pt plus 2pt minus 1pt}
\newcommand\afterCollaborationImgSpace{\vskip3pt plus 2pt minus 1pt}
\newcommand\afterAuthorSpace{\vskip5pt plus4pt minus4pt}
\newcommand\afterAffiliationSpace{\vskip3pt plus3pt}
\newcommand\afterEmailSpace{\vskip16pt plus9pt minus10pt\filbreak}
\newcommand\afterXtumSpace{\par\bigskip}
\newcommand\afterAbstractSpace{\vskip16pt plus9pt minus13pt}
\newcommand\afterKeywordsSpace{\vskip16pt plus9pt minus13pt}
\newcommand\afterArxivSpace{\vskip3pt plus0.01fil minus10pt}
\newcommand\afterDedicatedSpace{\vskip0pt plus0.01fil}
\newcommand\afterTocSpace{\bigskip\medskip}
\newcommand\afterTocRuleSpace{\bigskip\bigskip}
\newlength{\affiliationsSep}
\newcommand\beforetochook{\pagestyle{myplain}\pagenumbering{roman}}

\DeclareFixedFont\trfont{OT1}{phv}{b}{sc}{11}

\renewcommand\maketitle{
\pagestyle{empty}
\thispagestyle{titlepage}
\setcounter{page}{0}
\noindent{\small\scshape\@fpheader}\@preprint\par
\afterLogoSpace
\if!\@subheader!\else\noindent{\trfont{\@subheader}}\fi
\afterSubheaderSpace
\if!\@proceeding!\else\noindent{\sc\@proceeding}\fi
\afterProceedingsSpace
{\LARGE\flushleft\sffamily\bfseries\@title\par}
\afterTitleSpace
\hrule height 1.5\p@%
\afterRuleSpace
\if!\@collaboration!\else
{\Large\bfseries\sffamily\raggedright\@collaboration}\par
\afterCollaborationSpace
\fi
\if!\@collaborationImg!\else
{\normalsize\bfseries\sffamily\raggedright\@collaborationImg}\par
\afterCollaborationImgSpace
\fi
{\bfseries\raggedright\sffamily\the\auth@toks\par}
\afterAuthorSpace
\ifaffil\begin{list}{}{%
\setlength{\leftmargin}{0.28cm}%
\setlength{\labelsep}{0pt}%
\setlength{\itemsep}{\affiliationsSep}%
\setlength{\topsep}{-\parskip}}
\itshape\small%
\the\affil@toks
\end{list}\fi
\afterAffiliationSpace
\ifemailadd 
\noindent\hspace{0.28cm}\begin{minipage}[l]{.9\textwidth}
\begin{flushleft}
\textit{E-mail:} \the\email@toks
\end{flushleft}
\end{minipage}
\else 
\PackageWarningNoLine{\jname}{E-mails are missing.\MessageBreak Plese use \protect\emailAdd\space macro to provide e-mails.}
\fi
\afterEmailSpace
\if!\@xtum!\else\noindent{\@xtum}\afterXtumSpace\fi
\if!\@abstract!\else\noindent{\renewcommand\baselinestretch{.9}\textsc{Abstract:}}\ \@abstract\afterAbstractSpace\fi
\if!\@keywords!\else\noindent{\textsc{Keywords:}} \@keywords\afterKeywordsSpace\fi
\if!\@arxivnumber!\else\noindent{\textsc{ArXiv ePrint:}} \href{http://arxiv.org/abs/\@arxivnumber}{\@arxivnumber}\afterArxivSpace\fi
\if!\@dedicated!\else\vbox{\small\it\raggedleft\@dedicated}\afterDedicatedSpace\fi
\ifnotoc\else
\iftoccontinuous\else\newpage\fi
\beforetochook\hrule
\tableofcontents
\afterTocSpace
\hrule
\afterTocRuleSpace
\fi
\setcounter{footnote}{0}
\pagestyle{myplain}\pagenumbering{arabic}
} 

\renewcommand{\baselinestretch}{1.1}\normalsize
\divide\@tempdima\baselineskip
\@tempcnta=\@tempdima
\skip\@mpfootins = \skip\footins
\renewcommand{\@dotsep}{10000}

\newcommand\ps@myplain{
\pagenumbering{arabic}
\renewcommand\@oddfoot{\hfill-- \thepage\ --\hfill}
\renewcommand\@oddhead{}}
\let\ps@plain=\ps@myplain

\newcommand\ps@titlepage{\renewcommand\@oddfoot{}\renewcommand\@oddhead{}}

\numberwithin{equation}{section}

\renewcommand\section{\@startsection{section}{1}{\z@}%
                                   {-3.5ex \@plus -1.3ex \@minus -.7ex}%
                                   {2.3ex \@plus.4ex \@minus .4ex}%
                                   {\normalfont\large\bfseries}}
\renewcommand\subsection{\@startsection{subsection}{2}{\z@}%
                                   {-2.3ex\@plus -1ex \@minus -.5ex}%
                                   {1.2ex \@plus .3ex \@minus .3ex}%
                                   {\normalfont\normalsize\bfseries}}
\renewcommand\subsubsection{\@startsection{subsubsection}{3}{\z@}%
                                   {-2.3ex\@plus -1ex \@minus -.5ex}%
                                   {1ex \@plus .2ex \@minus .2ex}%
                                   {\normalfont\normalsize\bfseries}}
\renewcommand\paragraph{\@startsection{paragraph}{4}{\z@}%
                                   {1.75ex \@plus1ex \@minus.2ex}%
                                   {-1em}%
                                   {\normalfont\normalsize\bfseries}}
\renewcommand\subparagraph{\@startsection{subparagraph}{5}{\parindent}%
                                   {1.75ex \@plus1ex \@minus .2ex}%
                                   {-1em}%
                                   {\normalfont\normalsize\bfseries}}

\def\fnum@figure{\textbf{\figurename\nobreakspace\thefigure}}
\def\fnum@table{\textbf{\tablename\nobreakspace\thetable}}

\long\def\@makecaption#1#2{%
  \vskip\abovecaptionskip
  \sbox\@tempboxa{\small #1. #2}%
  \ifdim \wd\@tempboxa >\hsize
    \small #1. #2\par
  \else
    \global \@minipagefalse
    \hb@xt@\hsize{\hfil\box\@tempboxa\hfil}%
  \fi
  \vskip\belowcaptionskip}

\renewenvironment{thebibliography}[1]{%
\begin{oldthebibliography}{#1}%
\small%
\raggedright%
\setlength{\itemsep}{5pt plus 0.2ex minus 0.05ex}%
}%
{%
\end{oldthebibliography}%
}

\makeatother

\usepackage[T1]{fontenc} 
\usepackage{romannum}

\begin{document} 


\title{{\boldmath Exactly solvable Gross-Pitaevskii type equations} \\  }

\author[a]{Yuan-Yuan Liu,}
\author[a,b,c,1]{Wen-Du Li,}\note{liwendu@tjnu.edu.cn.}
\author[a,2]{and Wu-Sheng Dai}\note{daiwusheng@tju.edu.cn.}


\affiliation[a]{Department of Physics, Tianjin University, Tianjin 300350, P.R. China}
\affiliation[b]{College of Physics and Materials Science, Tianjin Normal University, Tianjin 300387, PR China}

\affiliation[c]{Theoretical Physics Division, Chern Institute of Mathematics, Nankai University, Tianjin, 300071, P. R. China}







\abstract{
We suggest a method to construct exactly solvable Gross-Pitaevskii type
equations, especially the variable-coefficient high-order Gross-Pitaevskii
type equations. We show that there exists a relation between the
Gross-Pitaevskii type equations. The Gross-Pitaevskii equations connected by
the relation form a family. In the family one only needs to solve one equation
and other equations in the family can be solved by a transform. That is, one
can construct a series of exactly solvable Gross-Pitaevskii type equations
from one exactly solvable Gross-Pitaevskii type equation. As examples, we
consider the family of some special Gross-Pitaevskii type equations:\ the
nonlinear Schr\"{o}dinger equation, the quintic Gross-Pitaevskii equation, and
cubic-quintic Gross-Pitaevskii equation. We also construct the family of a
kind of generalized Gross-Pitaevskii type equation.

}




\maketitle 

\flushbottom

\section{Introduction}

The Gross-Pitaevskii (GP) equation has many applications in various branches
of physics. In Bose-Einstein condensation, the GP equation is used to describe
dilute Bose gases \cite{pitaevskii2016bose,griesmaier2005bose}, the BEC in
optical lattices \cite{niu2020bose,hu2020vortices}, and the spinor BEC
\cite{yin2017solitons,lm2016efficient}. Moreover, the dynamics of BEC is
studied by the time-dependent GP equation from first-principle
\cite{benedikter2015quantitative} and the limit of GP equation's emergence is
also discussed \cite{nam2016ground}. Some methods based on the GP equation,
e.g., the truncated Wigner method \cite{norrie2006quantum}, the positive-P
method \cite{hope2001quantum}, the mean-field theory, and the
Hartree-Fock-Bogoliubov method \cite{rogel2004methods} are used to describe
BEC. The influence of the inter-particle interaction to BEC is an important
problem \cite{dai2017explicit,dai2007upper}, and the GP type equation is
suitable for describing the condensate of the interacting Bose gas. The GP
equation is also used to describe the Josephson plasma oscillations
\cite{burchianti2017josephson}. In general relativity, the GP equation is used
to study the gravastar of black hole physics \cite{cunillera2018the} and the
black hole in the anti-de Sitter space \cite{biasi2017delayed}. The GP
equation is a nonlinear equation and is difficult to solve. Many studies are
devoted to solving the GP equation, such as stationary solutions
\cite{charalampidis2018computing}, numerical solutions
\cite{antoine2018on,vergez2016a,sataric2016hybrid}, analytical solutions
\cite{liu2017analytical}, and soliton solutions
\cite{su2016nonautonomous,liu2017dark,gravejat2015asymptotic,zakeri2018solitons}%
. Some methods for solving the GP equation are developed, such as the inverse
scattering method \cite{yu2019inverse}. Some solutions of the GP equation with
various external potentials are obtained, e.g., the harmonic-oscillator
potential \cite{bland2018probing}, the multi-well potential
\cite{guo2018properties}, the changed external trap \cite{pickl2015derivation}%
, the nonlinear lattice pseudopotential \cite{alfimov2019localized}, the
external magnetic field \cite{olgiati2017remarks}, and a sort of
parity-time-symmetric potentials \cite{yu2019inverse,barashenkov2016exactly}.
The problem of scattering is also studied \cite{guo2018scattering}. Moreover,
there are studies on the one-dimensional GP equation and its applications
\cite{astrakharchik2018dynamics,mateo2011gap,modugno2018effective,piroli2016local,reichert2019casimir}%
. In this paper, we suggest a method for constructing exactly solvable GP type
equations, especially for variable-coefficient GP type equations. The GP
equation is also important in studying the nonlinear solitary and periodic
waves in the condensate of a superfluid \cite{Tsuzuki1971} and the ultracold
Bose-condensed atomic vapors in mesoscopic waveguide structures
\cite{ernst2010transport}.

The time-independent GP equation is
\begin{equation}
\nabla^{2}\psi\left(  \mathbf{r}\right)  -\left[  U_{\text{eff}}\left(
\mathbf{r}\right)  +g\left\vert \psi\left(  \mathbf{r}\right)  \right\vert
^{2}\right]  \psi\left(  \mathbf{r}\right)  =0, \label{1}%
\end{equation}
where $g$ is the coupling constant and $U_{\text{eff}}\left(  \mathbf{r}%
\right)  =U\left(  \mathbf{r}\right)  -\mu$ with $U\left(  \mathbf{r}\right)
$ the external potential and $\mu$ the chemical potential. Here we take
$\hbar=1$ and the mass $2m=1$ for simplicity. The variable-coefficient GP
equation, also called the inhomogeneous GP equation, whose coefficients are
space-dependent is a kind of important GP type equations
\cite{belmonte-beitia2009localized,yin2018periodic,belmonte-beitia2007lie}. In
the following, we consider the time-independent GP type equation in a general
form,%
\begin{equation}
\nabla^{2}\psi\left(  \mathbf{r}\right)  -\left[  U_{\text{eff}}\left(
\mathbf{r}\right)  +g_{1}\left(  \mathbf{r}\right)  \left\vert \psi\left(
\mathbf{r}\right)  \right\vert +g_{2}\left(  \mathbf{r}\right)  \left\vert
\psi\left(  \mathbf{r}\right)  \right\vert ^{2}+\cdots+g_{n}\left(
\mathbf{r}\right)  \left\vert \psi\left(  \mathbf{r}\right)  \right\vert
^{n}\right]  \psi\left(  \mathbf{r}\right)  =0, \label{4}%
\end{equation}
which includes any power of the density $\left\vert \psi\left(  \mathbf{r}%
\right)  \right\vert $ and the coefficient depends on the spatial coordinate.
The GP equation, the nonlinear Schr\"{o}dinger equation, etc., are the special
cases of Eq. (\ref{4}). The GP type equation is used to describe the many-body
interaction in BEC\textbf{
\cite{tang2007solution,wang2011quantized,shin2011bose,michinel2012coherent,zhang2012exact,wang2012localized,luckins2018bose,kohler2002three}%
.}\textit{ }On the other hand, the variable-coefficient GP type equation has a
wide application in nonlinear optics. The nonlinear Schr\"{o}dinger equation
performs numerical analysis of the pulse mechanism of laser
\cite{chu2020mode,nishizawa2019investigation,cai2017state}. The
variable-coefficient nonlinear Schr\"{o}dinger equation describes the pulse in
inhomogeneous optical systems \cite{meradji2020chirped} and the interactions
between periodic optical solitons \cite{wang226interactions}. The
variable-coefficient cubic-quintic GP equation describes the brightlike and
darklike solitary wave solutions \cite{li2020equivalence}.

In this paper, we show that there exists a relation between the GP type
equations. If two GP type equations are connected by the relation, their
solutions will be connected by a transform. The GP type equations who are
connected by the relation form a family. In a family, once an equation is
solved, the solutions of other equations in the family can be obtained by the transform.

As examples, we consider some families of the GP type equation, the family of
the GP equation, the family of the nonlinear Schr\"{o}dinger equation, the
family of the quintic GP equation, and the family of the cubic-quintic GP
equation. These GP type equations belong to different families. The solutions
of these equations are known, so by the transform we can solve all the
equations in their families.

In section \ref{1D}, we consider the family of the GP type equation. In
section \ref{family}, we discuss some exactly solvable families, including the
nonlinear Schr\"{o}dinger equation, the quintic GP equation, and the
cubic-quintic GP equation. In section \ref{1Dgeneralized}, we consider the
family of the generalized Gross-Pitaevskii type equation. In section
\ref{3DGP}, we consider the family of the three-dimensional spherically
symmetric GP type equation. The conclusions are summarized in section
\ref{Conclusion}.

\section{The family of the Gross-Pitaevskii type equation \label{1D}}

In this section, we show that there exists a relation between the GP type
equations. All the GP type equations who are related by the relation form a family.

\subsection{The relation}

For the GP type equations we have the following relation.

\textit{Two one-dimensional GP type equations}%
\begin{align}
\frac{d^{2}\psi\left(  x\right)  }{dx^{2}}-\left[  U_{\text{eff}}\left(
x\right)  +g_{1}\left(  x\right)  \left\vert \psi\left(  x\right)  \right\vert
+g_{2}\left(  x\right)  \left\vert \psi\left(  x\right)  \right\vert
^{2}+\cdots+g_{n}\left(  x\right)  \left\vert \psi\left(  x\right)
\right\vert ^{n}\right]  \psi\left(  x\right)   &  =0,\label{4.1}\\
\frac{d^{2}\phi\left(  \xi\right)  }{d\xi^{2}}-\left[  V_{\text{eff}}\left(
\xi\right)  +G_{1}\left(  \xi\right)  \left\vert \phi\left(  \xi\right)
\right\vert +G_{2}\left(  \xi\right)  \left\vert \phi\left(  \xi\right)
\right\vert ^{2}+\cdots+G_{n}\left(  \xi\right)  \left\vert \phi\left(
\xi\right)  \right\vert ^{n}\right]  \phi\left(  \xi\right)   &  =0,
\label{4.2}%
\end{align}
\textit{if }$U_{\text{eff}}\left(  x\right)  $ \textit{and} $V_{\text{eff}%
}\left(  \xi\right)  $\textit{, }$g_{l}\left(  x\right)  $ \textit{and}
$G_{l}\left(  \xi\right)  $ ($l=1,\cdots,n$)\textit{ satisfy the relations }%
\begin{align}
\sigma\left(  x^{2}U_{\text{eff}}\left(  x\right)  +\frac{1}{4}\right)   &
=\sigma^{-1}\left(  \xi^{2}V_{\text{eff}}\left(  \xi\right)  +\frac{1}%
{4}\right)  ,\label{4.3}\\
\sigma x^{\left(  l+4\right)  /2}g_{l}\left(  x\right)   &  =\sigma
^{-1}\left\vert \xi\right\vert ^{\left(  l+4\right)  /2}G_{l}\left(
\xi\right)  ,\text{ \ \ }l=1,\cdots,n, \label{4.5}%
\end{align}
\textit{their solutions are related by the transform: }%
\begin{align}
x  &  \leftrightarrow\left\vert \xi\right\vert ^{\sigma},\label{4.6}\\
\psi\left(  x\right)   &  \leftrightarrow\left\vert \xi\right\vert ^{^{\left(
\sigma-1\right)  /2}}\phi\left(  \xi\right)  . \label{4.7}%
\end{align}
\textit{Here }$\sigma$\textit{ is} \textit{a constant chosen arbitrarily.}

This result can be verified directly. Substituting the transforms (\ref{4.6})
and (\ref{4.7}) into the GP type equation (\ref{4.1}) gives%
\begin{align}
&  \frac{d^{2}\phi\left(  \xi\right)  }{d\xi^{2}}-\sigma^{2}\left[
\frac{1-\sigma^{-2}}{4\xi^{2}}+\left\vert \xi\right\vert ^{2\left(
\sigma-1\right)  }U_{\text{eff}}\left(  \left\vert \xi\right\vert ^{\sigma
}\right)  +\left\vert \xi\right\vert ^{\left(  5/2\right)  \left(
\sigma-1\right)  }g_{1}\left(  \left\vert \xi\right\vert ^{\sigma}\right)
\left\vert \phi\left(  \xi\right)  \right\vert \right. \nonumber\\
&  \left.  +\left\vert \xi\right\vert ^{3\left(  \sigma-1\right)  }%
g_{2}\left(  \left\vert \xi\right\vert ^{\sigma}\right)  \left\vert
\phi\left(  \xi\right)  \right\vert ^{2}+...+\left\vert \xi\right\vert
^{\left(  \sigma-1\right)  \left(  n+4\right)  /2}g_{n}\left(  \left\vert
\xi\right\vert ^{\sigma}\right)  \left\vert \phi\left(  \xi\right)
\right\vert ^{n}\right]  \phi\left(  \xi\right)  =0.
\end{align}
This is just the one-dimensional GP type equation (\ref{4.2}) with%
\begin{align}
V_{\text{eff}}\left(  \xi\right)   &  =\sigma^{2}\left[  \frac{1-\sigma^{-2}%
}{4\xi^{2}}+\left\vert \xi\right\vert ^{2\left(  \sigma-1\right)
}U_{\text{eff}}\left(  \left\vert \xi\right\vert ^{\sigma}\right)  \right]
,\nonumber\\
G_{l}\left(  \xi\right)   &  =\sigma^{2}\left\vert \xi\right\vert ^{\left(
\sigma-1\right)  \left(  l+4\right)  /2}g_{l}\left(  \left\vert \xi\right\vert
^{\sigma}\right)  ,\text{ }l=1,\cdots,n.
\end{align}
This proves the relations (\ref{4.3})-(\ref{4.7}).

\subsection{The family}

In the relation (\ref{4.3}) there is a constant $\sigma$. The constant
$\sigma$ can be chosen arbitrarily and different choices of $\sigma$ give
different transforms. This means that one GP type equation relates infinite
number of GP type equations through the relations (\ref{4.3})-(\ref{4.7}) with
different $\sigma$. The GP type equations who are related by a relation with
different $\sigma$ form a family. The family members are labeled by $\sigma$.
In a family, we only need to solve one equation and the solution of other
family members can be obtain directly by the transforms (\ref{4.6}) and
(\ref{4.7}).

In a GP type equation family, the family members are connected by a transform
with a transform parameter $\sigma$. This implies that there exists an
algebraic structure.

\subsection{The fixed point}

In a family, all family members are connected by a transform. This transform
has two fixed points.

The transforms (\ref{4.6}) and (\ref{4.7}) give%
\begin{equation}
\phi\left(  \xi\right)  =\left\vert \xi\right\vert ^{\left(  1-\sigma\right)
/2}\psi\left(  \left\vert \xi\right\vert ^{\sigma}\right)  .
\end{equation}
It can be seen directly that the points%
\begin{equation}
\left(  1,\psi\left(  1\right)  \right)  \text{ \ and }\left(  -1,\psi\left(
1\right)  \right)
\end{equation}
are fixed points in the transform. That is, in a family, all family members
pass through these two points.

\subsection{The family of the Gross-Pitaevskii equation \label{GPeqf}}

The GP equation is the most important special case of the GP type equation
(\ref{4}), which has only the $\left\vert \psi\left(  x\right)  \right\vert
^{2}$ term and a constant coefficient $g_{2}\left(  x\right)  =g$:%
\begin{equation}
\frac{d^{2}\psi\left(  x\right)  }{dx^{2}}-\left[  U_{\text{eff}}\left(
x\right)  +g\left\vert \psi\left(  x\right)  \right\vert ^{2}\right]
\psi\left(  x\right)  =0. \label{2.1}%
\end{equation}
The family of the GP equation, in which GP equation is one of its family
member, by the relations (\ref{4.3}) and (\ref{4.5}), consists of the
following family members:
\begin{equation}
\frac{d^{2}\phi\left(  \xi\right)  }{d\xi^{2}}-\left[  V_{\text{eff}}\left(
\xi\right)  +G\left(  \xi\right)  \left\vert \phi\left(  \xi\right)
\right\vert ^{2}\right]  \phi\left(  \xi\right)  =0 \label{gGPdual1D}%
\end{equation}
with\textit{ }%
\begin{align}
V_{\text{eff}}\left(  \xi\right)   &  =\sigma^{2}\left[  \frac{1-\sigma^{-2}%
}{4\xi^{2}}+\left\vert \xi\right\vert ^{2\left(  \sigma-1\right)
}U_{\text{eff}}\left(  \left\vert \xi\right\vert ^{\sigma}\right)  \right]
,\nonumber\\
G\left(  \xi\right)   &  =g\sigma^{2}\left\vert \xi\right\vert ^{3\left(
\sigma-1\right)  }. \label{GPCnormal}%
\end{align}
The family members are labeled by the parameter $\sigma$.

The solutions of family members are connected by the transforms (\ref{4.6})
and (\ref{4.7}):%
\begin{equation}
\phi\left(  \xi\right)  =\left\vert \xi\right\vert ^{\left(  1-\sigma\right)
/2}\psi\left(  \left\vert \xi\right\vert ^{\sigma}\right)  .
\end{equation}

\section{The exactly solvable family: examples \label{family}}

In the present paper, we consider the GP type equation in a general form. The
general GP type equation contains any power of the wave function with
space-dependent coefficients. Here we discuss the correspondence between the
general GP type equation and the BEC system.

Under the frame of the mean field theory, the interaction between two
particles is proportional to the particle number density $\left\vert
\psi\right\vert ^{2}$. Therefore, the term $\left\vert \psi\right\vert
^{2}\psi$ describes the two-body interaction, the term $\left\vert
\psi\right\vert ^{4}\psi$ describes the three-body interaction, and so on
\cite{tang2007solution,wang2011quantized,shin2011bose,michinel2012coherent,zhang2012exact,wang2012localized,luckins2018bose,kohler2002three}%
. That is, the even-power of $\left\vert \psi\right\vert $ describes the
many-body interaction. Moreover, the odd-power term is also used to describe
the effect beyond the mean field treatment \cite{Ilg2018,Debnath2020}.

The coupling parameter $g$ is proportional to the $s$-wave scattering length
of inter-atomic scattering in BEC
\cite{pitajevskij2003bose,kengne2020spatiotemporal} and the scattering length
can be determined by the Feshbach resonance \cite{michinel2012coherent}. In
the BEC experiment, the coupling parameter can be controlled by the Feshbach
resonance which changes the scattering length
\cite{pinsker2013nonlinear,vilchynskyy2013nature}. For example, the scattering
length can be changed by the magnetic field
\cite{feshbach1967intermediate,chin2010feshbach}. The spatial modulation leads
to a space-dependent coupling parameter and the temporal modulation leads to a
time-dependent coupling parameter
\cite{kengne2020spatiotemporal,pinsker2013nonlinear,combescot2003feshbach}. In
the present paper, we consider the GP type equations with space-dependent
coupling parameter.

For two-body interactions, the coupling parameter of the term $\left\vert
\psi\right\vert ^{2}\psi$ is of the magnitude
\cite{wang2011quantized,wang2012localized}%
\begin{equation}
g\sim\frac{4\pi\hbar^{2}}{m}a_{s},
\end{equation}
where $a_{s}$ is the $s$-wave scattering length. If only consider two-body
interactions, it should satisfy $\sqrt{na_{s}^{3}}\ll1$ with $n$ the particle
number density.

In the case of high densities, the three-body interaction becomes important
and the two-body description is no longer effective \cite{wang2012localized}.
Moreover, the coupling parameter of the three-body interaction, i.e., the
coefficient of the term $\left\vert \psi\right\vert ^{4}\psi$, is of the
magnitude \cite{wang2012localized}%
\begin{equation}
g_{3}=\frac{12\pi\hbar^{2}a_{s}^{4}}{m}\left(  d_{1}+d_{2}\right)  \tan\left(
s_{0}\ln\left(  \frac{a_{s}}{a_{0}}+\frac{\pi}{2}\right)  \right)  \label{g3}%
\end{equation}
which is proportional to $a_{s}^{4}$, so when the scattering length is large,
the three-body interaction\ needs to be taken into account. The coupling
parameter $g_{3}$ may be a complex number. For example, for the $^{87}$Rb
condensate the real part of $g_{3}/\hbar$ is about $10^{-26}\sim10^{-27}%
$cm$^{6}$s$^{-1}$
\cite{wang2012localized,bulgac2002dilute,pieri2003derivation} and the
imaginary part is about $10^{-30}$cm$^{6}$s$^{-1}$ and is ignorable
\cite{tolra2004observation}. The other parameters in Eq. (\ref{g3}) can be
determined numerically \cite{bulgac2002dilute,braaten2002dilute}. 

In this section, we consider some exactly solvable families, the families of
the nonlinear Schr\"{o}dinger equation, the quintic GP equation, and the
cubic-quintic GP equation.

\subsection{The family of the nonlinear Schr\"{o}dinger equation
\label{NSLeqf}}

The stationary nonlinear Schr\"{o}dinger equation
\cite{gnutzmann2011stationary}%
\begin{equation}
\frac{d^{2}\psi\left(  x\right)  }{dx^{2}}+\left[  E-g\left\vert \psi\left(
x\right)  \right\vert ^{2}\right]  \psi\left(  x\right)  =0 \label{1.1}%
\end{equation}
is a special case of the GP equation (\ref{2.1}) with a vanishing external
potential and the chemical potential $\mu$ replaced by the energy $E$.

The family of the nonlinear Schr\"{o}dinger equation, by the relations
(\ref{4.3}) and (\ref{4.5}), consists of the following family members:%
\begin{equation}
\frac{d^{2}\phi\left(  \xi\right)  }{d\xi^{2}}-\left[  V_{\text{eff}}\left(
\xi\right)  +G\left(  \xi\right)  \left\vert \phi\left(  \xi\right)
\right\vert ^{2}\right]  \phi\left(  \xi\right)  =0 \label{1.2}%
\end{equation}
with\textit{ }%
\begin{align}
V_{\text{eff}}\left(  \xi\right)   &  =\sigma^{2}\left[  \frac{1-\sigma^{-2}%
}{4\xi^{2}}-\left\vert \xi\right\vert ^{2\left(  \sigma-1\right)  }E\right]
,\nonumber\\
G\left(  \xi\right)   &  =g\sigma^{2}\left\vert \xi\right\vert ^{3\left(
\sigma-1\right)  }, \label{NSEG}%
\end{align}
where $V_{\text{eff}}\left(  \xi\right)  =V\left(  \xi\right)  -\mathcal{E}$.
The solution of Eq. (\ref{1.2}) by the transforms (\ref{4.6}) and (\ref{4.7})
is%
\begin{equation}
\phi\left(  \xi\right)  =\left\vert \xi\right\vert ^{\left(  1-\sigma\right)
/2}\psi\left(  \left\vert \xi\right\vert ^{\sigma}\right)  . \label{NLSTr2}%
\end{equation}

It can be checked that the nonlinear Schr\"{o}dinger equation (\ref{1.1}) has
a solution%
\begin{equation}
\psi\left(  x\right)  =\sqrt{\frac{E}{g}}\tanh\left(  \sqrt{\frac{E}{2}%
}\left(  x+b\right)  \right)  ,
\end{equation}
where $b$ is a constant.

Then the solution of the family member, Eq. (\ref{1.2}), by the transform
(\ref{NLSTr2}) is%
\begin{equation}
\phi\left(  \xi\right)  =\left\vert \sigma\right\vert \left\vert
\xi\right\vert ^{\sigma-1}\sqrt{\frac{E}{G\left(  \xi\right)  }}\tanh\left(
\sqrt{\frac{E}{2}}\left(  \left\vert \xi\right\vert ^{\sigma}+b\right)
\right)  .
\end{equation}

\begin{figure}[ptb]
\centering\includegraphics[width=1.0\textwidth]{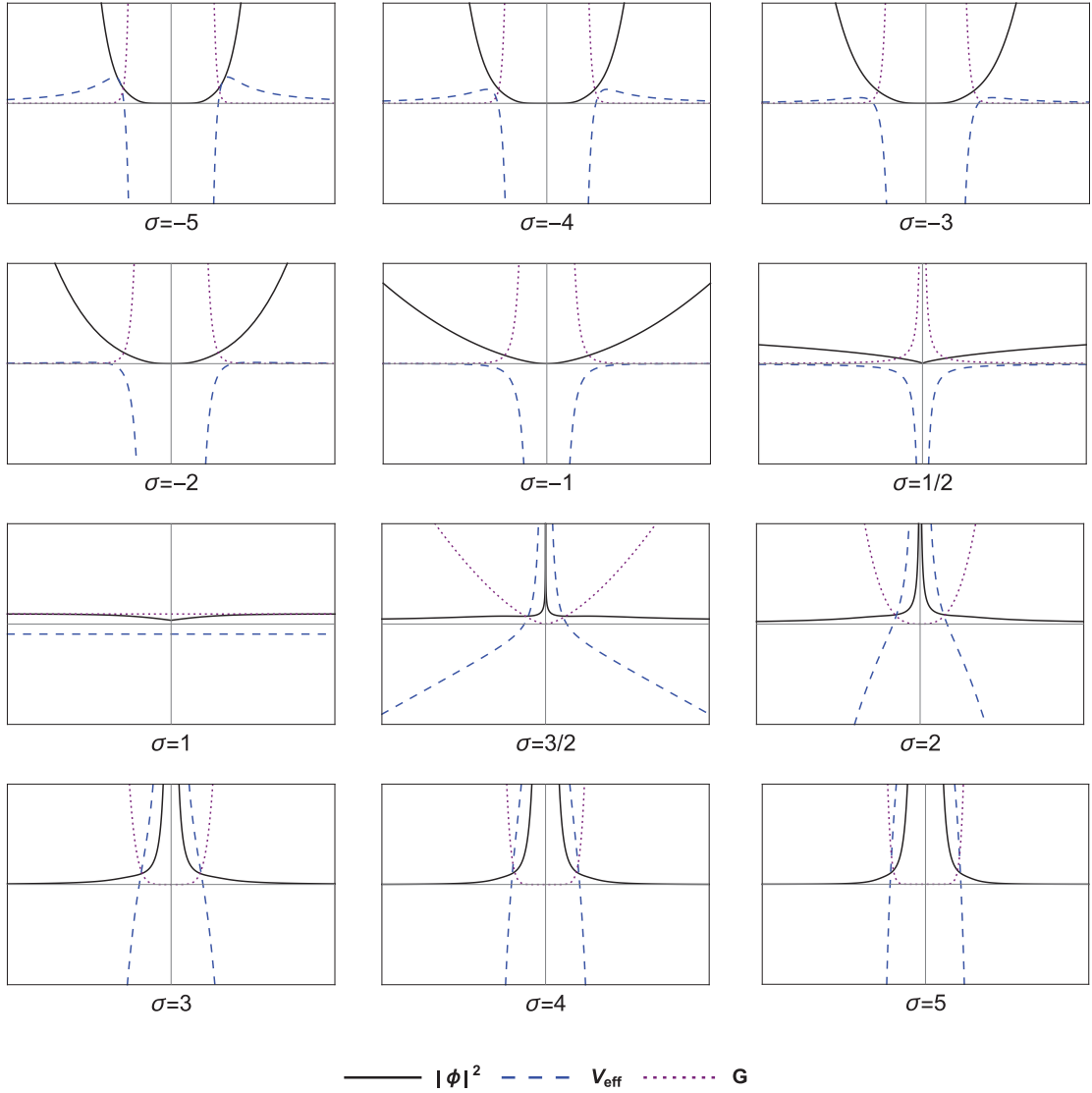}\newline\caption{The
family members of the nonlinear Schr\"{o}dinger equation with various value of
$\sigma$. }%
\label{figNLSnew}%
\end{figure}

The family members of the nonlinear Schr\"{o}dinger equation with various
value of $\sigma$ are shown in Fig. (\ref{figNLSnew}).

For a special case of the stationary nonlinear Schr\"{o}dinger equation
(\ref{1.1})%
\begin{equation}
\frac{d^{2}\psi\left(  x\right)  }{dx^{2}}+\left[  2\mu+2\left\vert
\psi\left(  x\right)  \right\vert ^{2}\right]  \psi\left(  x\right)
=0,\label{NewNLS}%
\end{equation}
Ref. \cite{witthaut2005bound} provides an exact bright soliton solution:%
\begin{equation}
\psi\left(  x\right)  =\sqrt{-2\mu}\operatorname{sech}\left[  \sqrt{-2\mu
}\left(  x-x_{0}\right)  \right]  .
\end{equation}

The family of the nonlinear Schr\"{o}dinger equation (\ref{NewNLS}), by the
relations (\ref{4.3}) and (\ref{4.5}), consists of the following family
members:%
\begin{equation}
\frac{d^{2}\phi\left(  \xi\right)  }{d\xi^{2}}-\left[  V_{eff}\left(
\xi\right)  +G_{2}\left(  \xi\right)  \left\vert \phi\left(  \xi\right)
\right\vert ^{2}\right]  \phi\left(  \xi\right)  =0,\label{fms}%
\end{equation}
where%
\begin{align}
V_{eff}\left(  \xi\right)   &  =\sigma^{2}\left[  \frac{1-\sigma^{-2}}%
{4\xi^{2}}-2\mu\left\vert \xi\right\vert ^{2\left(  \sigma-1\right)  }\right]
,\nonumber\\
G_{2}\left(  \xi\right)   &  =-2\sigma^{2}\left\vert \xi\right\vert ^{3\left(
\sigma-1\right)  }.
\end{align}
The solution of the family members (\ref{fms}) is
\begin{align}
\phi\left(  \xi\right)   &  =\left\vert \xi\right\vert ^{\left(
1-\sigma\right)  /2}\psi\left(  \left\vert \xi\right\vert ^{\sigma}\right)
\nonumber\\
&  =\left\vert \xi\right\vert ^{\left(  1-\sigma\right)  /2}\sqrt{-2\mu
}\operatorname{sech}\left[  \sqrt{-2\mu}\left(  \left\vert \xi\right\vert
^{\sigma}-\xi_{0}\right)  \right]  .
\end{align}

\subsection{The family of the quintic Gross-Pitaevskii equation \label{FqGP}}

The quintic GP equation describes BEC when the interaction between the atoms
is moderate or strong \cite{luckins2018bose}.

The quintic GP equation
\begin{equation}
\frac{d^{2}\psi\left(  x\right)  }{dx^{2}}-\left[  1+g_{4}\left\vert
\psi\left(  x\right)  \right\vert ^{4}\right]  \psi\left(  x\right)  =0
\label{qGP}%
\end{equation}
has a solution \cite{luckins2018bose}%
\begin{equation}
\psi\left(  x\right)  =\left(  \frac{3}{-g_{4}}\right)  ^{1/4}\frac{1}%
{\sqrt{\cosh\left(  2x\right)  }},
\end{equation}
where $g_{4}$ is a negative constant.

The family of the quintic GP equation, by the relations (\ref{4.3}) and
(\ref{4.5}), consists of the family members:%
\begin{equation}
\frac{d^{2}\phi\left(  \xi\right)  }{d\xi^{2}}-\left[  V_{\text{eff}}\left(
\xi\right)  +G_{4}\left(  \xi\right)  \left\vert \phi\left(  \xi\right)
\right\vert ^{4}\right]  \phi\left(  \xi\right)  =0, \label{dualqGP}%
\end{equation}
with%
\begin{align}
V_{\text{eff}}\left(  \xi\right)   &  =\sigma^{2}\left[  \frac{1-\sigma^{-2}%
}{4\xi^{2}}+\left\vert \xi\right\vert ^{2\left(  \sigma-1\right)  }\right]
,\nonumber\\
G_{4}\left(  \xi\right)   &  =\sigma^{2}\left\vert \xi\right\vert ^{4\left(
\sigma-1\right)  }g_{4}.
\end{align}

The solution of Eq. (\ref{dualqGP}) by the transforms (\ref{4.6}) and
(\ref{4.7}) is%
\begin{equation}
\phi\left(  \xi\right)  =\left\vert \sigma\right\vert ^{1/2}\left\vert
\xi\right\vert ^{\left(  \sigma-1\right)  /2}\left(  \frac{3}{-G_{4}\left(
\xi\right)  }\right)  ^{1/4}\frac{1}{\sqrt{\cosh\left(  2\left\vert
\xi\right\vert ^{\sigma}\right)  }}.
\end{equation}

\subsection{The family of the cubic-quintic Gross-Pitaevskii equation
\label{FcqGP}}

The cubic-quintic GP equation which describes BEC considers two-particle and
three-particle interactions \cite{luckins2018bose,kohler2002three}.

The cubic-quintic GP equation
\begin{equation}
\frac{d^{2}\psi\left(  x\right)  }{dx^{2}}-\left[  1+g_{2}\left\vert
\psi\left(  x\right)  \right\vert ^{2}+g_{4}\left\vert \psi\left(  x\right)
\right\vert ^{4}\right]  \psi\left(  x\right)  =0 \label{cqGP}%
\end{equation}
has a solution \cite{luckins2018bose}%
\begin{equation}
\psi\left(  x\right)  =\frac{2}{\sqrt{\sqrt{g_{2}^{2}-\frac{16}{3}g_{4}}%
\cosh\left(  2x\right)  -g_{2}}},
\end{equation}
where $g_{2}$ and $g_{4}$ are negative constants.

The family of the cubic-quintic GP equation, by the relations (\ref{4.3}) and
(\ref{4.5}), consists of the family members:%
\begin{equation}
\frac{d^{2}\phi\left(  \xi\right)  }{d\xi^{2}}-\left[  V_{\text{eff}}\left(
\xi\right)  +G_{2}\left(  \xi\right)  \left\vert \phi\left(  \xi\right)
\right\vert ^{2}+G_{4}\left(  \xi\right)  \left\vert \phi\left(  \xi\right)
\right\vert ^{4}\right]  \phi\left(  \xi\right)  =0 \label{dualcqGP}%
\end{equation}
with%
\begin{align}
V_{\text{eff}}\left(  \xi\right)   &  =\sigma^{2}\left[  \frac{1-\sigma^{-2}%
}{4\xi^{2}}+\left\vert \xi\right\vert ^{2\left(  \sigma-1\right)  }\right]
,\nonumber\\
G_{2}\left(  \xi\right)   &  =\sigma^{2}\left\vert \xi\right\vert ^{3\left(
\sigma-1\right)  }g_{2},\nonumber\\
G_{4}\left(  \xi\right)   &  =\sigma^{2}\left\vert \xi\right\vert ^{4\left(
\sigma-1\right)  }g_{4}.
\end{align}
The solution of Eq. (\ref{dualcqGP}) by the transforms (\ref{4.6}) and
(\ref{4.7}) is
\begin{equation}
\phi\left(  \xi\right)  =\left\vert \sigma\right\vert ^{1/2}\left\vert
\xi\right\vert ^{\left(  \sigma-1\right)  /2}\frac{2}{\sqrt{\sqrt{\left[
\left\vert \sigma\right\vert ^{-1}\left\vert \xi\right\vert ^{1-\sigma}%
G_{2}\left(  \xi\right)  \right]  ^{2}-\frac{16}{3}G_{4}\left(  \xi\right)
}\cosh\left(  2\left\vert \xi\right\vert ^{\sigma}\right)  -\left\vert
\sigma\right\vert ^{-1}\left\vert \xi\right\vert ^{1-\sigma}G_{2}\left(
\xi\right)  }}.
\end{equation}

\section{The family of the generalized Gross-Pitaevskii type equation
\label{1Dgeneralized}}

For academic interest, we consider a generalized Gross-Pitaevskii type
equation and its family. By the generalized Gross-Pitaevskii type equation we
mean that $\left\vert \psi\left(  x\right)  \right\vert ^{n}\psi\left(
x\right)  $ in the Gross-Pitaevskii type equation is replaced by $\psi\left(
x\right)  ^{n+1}$.

\textit{Two one-dimensional generalized GP type equations}%
\begin{align}
\frac{d^{2}\psi\left(  x\right)  }{dx^{2}}-\left[  U_{\text{eff}}\left(
x\right)  +g_{1}\left(  x\right)  \psi\left(  x\right)  +g_{2}\left(
x\right)  \psi^{2}\left(  x\right)  +\cdots+g_{n}\left(  x\right)  \psi
^{n}\left(  x\right)  \right]  \psi\left(  x\right)   &  =0,\label{gGP}\\
\frac{d^{2}\phi\left(  \xi\right)  }{d\xi^{2}}-\left[  V_{\text{eff}}\left(
\xi\right)  +G_{1}\left(  \xi\right)  \psi\left(  \xi\right)  +G_{2}\left(
\xi\right)  \phi^{2}\left(  \xi\right)  +\cdots+G_{n}\left(  \xi\right)
\phi^{n}\left(  \xi\right)  \right]  \phi\left(  \xi\right)   &  =0,
\label{dgGP}%
\end{align}
\textit{if }$U_{\text{eff}}\left(  x\right)  $ \textit{and} $V_{\text{eff}%
}\left(  \xi\right)  $\textit{, }$g_{l}\left(  x\right)  $ \textit{and}
$G_{l}\left(  \xi\right)  $ ($l=1,\cdots,n$)\textit{ satisfy the relations}%
\begin{align}
\sigma\left(  x^{2}U_{\text{eff}}\left(  x\right)  +\frac{1}{4}\right)   &
=\sigma^{-1}\left(  \xi^{2}V_{\text{eff}}\left(  \xi\right)  +\frac{1}%
{4}\right)  ,\nonumber\\
\sigma x^{\left(  l+4\right)  /2}g_{l}\left(  x\right)   &  =\sigma^{-1}%
\xi^{\left(  l+4\right)  /2}G_{l}\left(  \xi\right)  ,\text{ \ \ }%
l=1,\cdots,n, \label{G2}%
\end{align}
\textit{their solutions are related by the transform:}%
\begin{align}
x  &  \leftrightarrow\xi^{\sigma},\nonumber\\
\psi\left(  x\right)   &  \leftrightarrow\xi^{\left(  \sigma-1\right)  /2}%
\phi\left(  \xi\right)  .
\end{align}
\textit{Here }$\sigma$\textit{ is} \textit{a constant chosen arbitrarily.}

This result can be verified directly by the same procedure as for the GP type equation.

The fixed points by the transform (\ref{G2}), i.e.,%
\begin{equation}
\phi\left(  \xi\right)  =\xi^{\left(  1-\sigma\right)  /2}\psi\left(
\xi^{\sigma}\right)  .
\end{equation}
are
\begin{equation}
\left(  1,\psi\left(  1\right)  \right)  \text{ \ and \ }\left(
-1,\psi\left(  -1\right)  \right)  .
\end{equation}
In a family, all family members pass through these two points.

The generalized GP type equation (\ref{gGP}) is a Li\'{e}nard equation
\cite{yang2008exact,khan2013bell,solomon1977differential} with space-dependent
coefficients. The result obtained here can be used to consider the family of
the Li\'{e}nard equation.

\section{The three-dimensional spherically symmetric Gross-Pitaevskii type
equation \label{3DGP}}

The three dimensional spherically symmetric GP type equation also has the
similar relation.

\textit{Two three-dimensional radial GP type equations}%
\begin{align}
\frac{d^{2}\psi_{l}\left(  r\right)  }{dr^{2}}-\left[  U_{\text{eff}}\left(
r\right)  +\frac{l\left(  l+1\right)  }{r^{2}}+g\left(  r\right)  \frac
{1}{r^{2}}\left\vert \psi_{l}\left(  r\right)  \right\vert ^{2}\right]
\psi_{l}\left(  r\right)   &  =0,\label{3DGPtypeU}\\
\frac{d^{2}\phi_{\ell}\left(  \rho\right)  }{d\rho^{2}}-\left[  V_{\text{eff}%
}\left(  \rho\right)  +\frac{\ell\left(  \ell+1\right)  }{\rho^{2}}+G\left(
\rho\right)  \frac{1}{\rho^{2}}\left\vert \phi_{\ell}\left(  \rho\right)
\right\vert ^{2}\right]  \phi_{\ell}\left(  \rho\right)   &  =0,
\label{3DGPtypeV}%
\end{align}
\textit{where }$l$\textit{ and }$\ell$\textit{ are angular quantum numbers, if
}$U_{\text{eff}}\left(  r\right)  $ \textit{and} $V_{\text{eff}}\left(
\rho\right)  $\textit{, }$g\left(  r\right)  $ and $G\left(  \rho\right)
$\textit{ satisfy the relations}%
\begin{align}
\sigma r^{2}U_{\text{eff}}\left(  r\right)   &  =\sigma^{-1}\rho
^{2}V_{\text{eff}}\left(  \rho\right)  ,\nonumber\\
\sigma rg\left(  r\right)   &  =\sigma^{-1}\rho G\left(  \rho\right)  ,
\label{3.4}%
\end{align}
\textit{their solutions are related by the transform:}%
\begin{align}
r  &  \leftrightarrow\rho^{\sigma},\nonumber\\
\psi_{l}\left(  r\right)   &  \leftrightarrow\rho^{\left(  \sigma-1\right)
/2}\phi_{\ell}\left(  \rho\right)  ,\nonumber\\
l+\frac{1}{2}  &  \leftrightarrow\sigma^{-1}\left(  \ell+\frac{1}{2}\right)  .
\label{3.7}%
\end{align}
\textit{Here }$\sigma$ \textit{is a constant chosen arbitrarily.}

This result can be verified directly. Substituting the transform (\ref{3.7})
into Eq. (\ref{3DGPtypeU}) gives%
\begin{equation}
\frac{d^{2}\phi_{\ell}\left(  \rho\right)  }{d\rho^{2}}-\left[  \sigma^{2}%
\rho^{2\left(  \sigma-1\right)  }U_{\text{eff}}\left(  \rho^{\sigma}\right)
+\frac{\ell\left(  \ell+1\right)  }{\rho^{2}}+g\left(  \rho^{\sigma}\right)
\sigma^{2}\rho^{\sigma-1}\frac{1}{\rho^{2}}\left\vert \phi_{\ell}\left(
\rho\right)  \right\vert ^{2}\right]  \phi_{\ell}\left(  \rho\right)  =0,
\end{equation}
This is just the three-dimensional radial GP type equation (\ref{3DGPtypeV})
with%
\begin{align}
V_{\text{eff}}\left(  \rho\right)   &  =\sigma^{2}\rho^{2\left(
\sigma-1\right)  }U_{\text{eff}}\left(  \rho^{\sigma}\right)  ,\nonumber\\
G\left(  \rho\right)   &  =g\left(  \rho^{\sigma}\right)  \sigma^{2}%
\rho^{\sigma-1}.
\end{align}
This proves the relations (\ref{3.4}) and (\ref{3.7}).

\section{Conclusion \label{Conclusion}}

We show that there exist families of the GP type equations. The GP type
equations in a family are related by a transform. In a family, so long as one
family member is solved, all family members are solved by the transform. The
GP type equation is difficult to solve. The method presented in the paper
provides an approach to construct exactly solvable GP type equations.

As examples, we consider the family of some special GP type equations: the
nonlinear Schr\"{o}dinger equation, the quintic GP equation, and the
cubic-quintic GP equation.

We also consider family of the generalized GP type equation. The result of the
generalized GP type equation inspires us to consider the family of the
Li\'{e}nard equation in the future work.

For three-dimensional cases, we consider the family of the three-dimensional
spherically symmetric GP type equation.

\bigskip

\acknowledgments

We are very indebted to Dr G. Zeitrauman for his encouragement. This work is supported in part by Special Funds for Theoretical Physics Research Program of the National Natural Science Foundation of China under Grant No. 11947124 and NSF of China under Grant No. 11575125 and No. 11675119.







\providecommand{\href}[2]{#2}\begingroup\raggedright\endgroup



\end{document}